\documentclass[webpdf,modern,large]{oup-authoring-template}

\graphicspath{{figures/}}
\usepackage{microtype}
\usepackage{stfloats}
\hypersetup{colorlinks=true, citecolor=blue, linkcolor=blue, urlcolor=blue}
\setcitestyle{aysep={},citesep={,}}
\makeatletter
\def\secsize{\sffamily\fontsize{11bp}{13bp}\selectfont\bfseries}
\def\subsecsize{\sffamily\fontsize{10bp}{12bp}\selectfont\bfseries}
\def\subsubsecsize{\sffamily\fontsize{9.5bp}{11.5bp}\selectfont\bfseries\itshape}
\makeatother

\newcommand{\ddG}{\ensuremath{\Delta\Delta G}}

\newcommand{\kcalmol}{\ensuremath{\mathrm{kcal\,mol^{-1}}}}
\newcommand{\alttext}[1]{\par\smallskip\noindent\footnotesize\textbf{Alt text:} #1\par} % printed alt text, required by OUP at submission

\begin{document}

\journaltitle{Bioinformatics}
\DOI{}
\copyrightyear{}
\pubyear{}
\vol{}
\issue{}
\access{}
\appnotes{Original Paper}
\firstpage{1}

\subtitle{Structural Bioinformatics}
\title[CIR-DDG]{CIR-DDG: backbone-agnostic residual correction of antibody--antigen affinity changes with explicit cross-chain geometry}

% AUTHOR ACTION: replace all author and affiliation placeholders.
\author[1]{Weilun Yu}
\author[3]{Zhiheng Zou}
\author[4]{Yonggui Huang}
\author[5]{Honggang Qi}
\author[1]{Geguang Pu}
\author[1]{Yu Xiao}
\author[1]{Gang Xu}
\author[1]{Jiangtao Wang}
\author[1,2,$\ast$]{Xi Chen}
\address[1]{\orgdiv{Software Engineering Institute}, \orgname{East China Normal University}, \orgaddress{\street{Shanghai}, \postcode{200062}, \country{China}}}
\address[2]{\orgdiv{Shanghai Key Laboratory of Multidimensional Information Processing},\orgname{East China Normal University}, \orgaddress{\street{Shanghai}, \postcode{200062}, \country{China}}}
\address[3]{\orgname{Jiangxi University of Science and Technology}, \orgaddress{\street{Ganzhou}, \postcode{341099}, \country{China}}}
\address[4]{\orgname{Peking University}, \orgaddress{\street{Beijing}, \postcode{100871}, \country{China}}}
\address[5]{\orgdiv{School of Computer Science and Technology}, \orgname{University of Chinese Academy of Sciences}, \orgaddress{\street{Beijing}, \postcode{100049}, \country{China}}}

\corresp[$\ast$]{Corresponding author. \href{mailto:xchen@geo.ecnu.edu.cn}{xchen@geo.ecnu.edu.cn}}

\abstract{\mdseries\fontsize{8.5bp}{11.5bp}\selectfont
\textbf{Motivation:} Accurate prediction of mutation-induced protein--protein binding free-energy changes is important for antibody affinity maturation, yet scarce labels and complex interface geometry limit generalization. Heterogeneous predictors may process three-dimensional complexes without preserving the cross-chain signals most relevant to a mutation in their final scalar output.\\
\textbf{Results:} We introduce CIR-DDG, a lightweight residual adapter that combines a fixed base prediction with 22 interpretable descriptors of cross-chain distance, contact density and site--partner context. In complex-level five-fold evaluation on SKEMPI~2.0 measurements from 343 complexes, CIR-DDG improved all six tested backbones on antibody--antigen interface mutations: Spearman correlation increased by 0.0346--0.1296, while RMSE decreased by 0.0074--0.0408~\kcalmol. Cross-validated probing, equal-capacity controls and feature ablations support the complementarity of explicit geometry. On an independent SARS-CoV-2 RBD--ACE2 deep-mutational-scan benchmark of 3669 substitutions, the fold-specific adapters transferred without any retraining: the absolute interface Spearman correlation increased by 0.026--0.081 for all four evaluable backbones, showing that the learned geometric correction generalizes beyond SKEMPI thermodynamic measurements.\\
\textbf{Availability and implementation:} CIR-DDG is available at \url{https://github.com/ecnuabmlab/CIR-ddG}.}

% \keywords{antibody--antigen interaction, binding free-energy change, mutation effect prediction, interface geometry, residual learning, protein engineering}

\maketitle

\section{Introduction}

Antibodies recognize antigens through highly specific non-covalent interactions between complementarity-determining regions (CDRs) and antigen epitopes. Binding strength influences neutralization, selectivity and therapeutic efficacy, but affinity maturation requires searching a combinatorial sequence space in which beneficial substitutions are rare. Recent experimental studies \citep{makowski2022co, hie2024efficient} show that machine learning can focus this search, improve affinity with small screening budgets and generate diverse high-affinity libraries; they also expose trade-offs between affinity and non-specific binding that a single-objective score cannot capture. Experimental assays such as isothermal titration calorimetry, surface plasmon resonance and deep mutational scanning nevertheless remain limited by cost, throughput and system coverage. Computational prediction of the mutation-induced change in binding free energy, \ddG, is therefore an important component of rational antibody optimization and protein engineering. SKEMPI~2.0 \citep{jankauskaite2019skempi} provides a widely used structure--thermodynamics benchmark for this task, although its complexes, mutation types and measured effects are unevenly distributed.

Classical predictors established two complementary paradigms. Physics-based tools evaluate stability and interaction changes with empirical or all-atom energy functions, as in FoldX~5 \citep{delgado2019foldx} and Rosetta Flex~ddG \citep{kortemme2002simple, barlow2018flex}, but conformational sampling and local relaxation increase the cost of large mutation screens. A second line replaces or augments energy calculations with statistical and engineered interface representations: BeAtMuSiC \citep{dehouck2013beatmusic} combines coarse-grained statistical potentials, and mCSM \citep{pires2014mcsm} encodes atomic-distance patterns around the mutation. Their success established that compact local geometry and partner context can complement global physical scores, while also highlighting the dependence of classical machine learning on hand-designed features.

More recent predictors learn representations directly from sequence and structure. ESM-IF \citep{hsu2022learning} and ProteinMPNN \citep{dauparas2022robust} provide zero-shot scores from structure-conditioned sequence likelihoods; RDE-Network \citep{luo2023rotamer} and DiffAffinity \citep{liu2023predicting} learn side-chain conformational statistics; and Prompt-DDG \citep{wu2024learning} models mutation-centred microenvironments with hierarchical prompts. DDAffinity \citep{yu2024ddaffinity}, PPIformer \citep{bushuiev2024learning} and Pythia-PPI \citep{tao2025reliable} incorporate supervised geometric learning, interface pre-training or auxiliary stability data. Analyses of benchmark bias and antibody-specific data requirements \citep{tsishyn2024quantification, hummer2025investigating} show that apparent accuracy depends strongly on structural similarity, split construction and training-set diversity.

A parallel experimental development supplies orthogonal test grounds for these predictors. Yeast-display deep mutational scanning (DMS) measures binding effects for near-saturating single substitutions at one interface \citep{starr2022shifting}, providing thousands of quantitative measurements that are independent of thermodynamic databases. The SARS-CoV-2 spike receptor-binding domain (RBD) bound to human ACE2 has become a \emph{de facto} external case study: DDAffinity \citep{yu2024ddaffinity} and CATH-ddG \citep{yu2025cath} both report RBD benchmarks, albeit restricted to a few hundred substitutions at pre-selected sites. DMS data measure binding preferences rather than calorimetric free energies, and their sign and scale conventions differ from SKEMPI, so they probe the transfer of interface sensitivity rather than absolute thermodynamic calibration. We use a full-coverage RBD DMS cohort in exactly this role.

A complementary modelling perspective treats an existing predictor as a fixed prior and learns a small corrective module on its output, in the spirit of parameter-efficient adapter modules \citep{houlsby2019adapters} in transfer learning. Output-space correction has three practical attractions for \ddG{} prediction: it requires neither retraining nor access to backbone internals; it can be restricted to a defined cohort so that behaviour elsewhere is preserved by construction; and its small capacity acts as a regularizer under scarce labels. Whether a minimal adapter can recover interface information missed by structurally aware backbones is an empirical question, and answering it requires controls against trivial parameter-count effects.

The representations learned by the predictors reviewed above are valuable priors for \ddG, but their training objectives need not preserve a compact description of the geometry between a mutated interface residue and the opposing chain at the final prediction layer. The backbones evaluated here include zero-shot inverse-folding models, unsupervised side-chain models and end-to-end \ddG{} predictors, several of which do process three-dimensional complexes. We therefore do not assume that they ignore structure; we test the narrower hypothesis that their scalar outputs under-represent explicit cross-chain distances, contact counts and partner chemistry that remain predictive of antibody--antigen mutation effects.

We propose Cross-Chain Interaction Residual for \ddG{} (CIR-DDG), a backbone-agnostic module that adds an interface-specific correction to an existing prediction. CIR-DDG uses 22 physically interpretable geometric descriptors and an interface mask that leaves out-of-scope predictions unchanged. We evaluate the same adapter with six heterogeneous backbones under a shared complex-level split, then challenge it without retraining on an independent SARS-CoV-2 RBD--ACE2 deep-mutational-scan benchmark of 3669 substitutions. The study connects target-interface performance, scope conservation, information probing, equal-capacity and feature controls, external transfer and a held-out-complex case analysis. This progression goes beyond asking whether CIR-DDG improves prediction: it tests whether a small output-space adapter can recover interface information complementary to diverse backbone scores without erasing their established behaviour.

The main contributions are as follows. (i)~We formalize output-level interface under-encoding and quantify it with cross-validated ridge probes that separate the information a backbone exposes in its scalar score from the information explicit geometry still adds. (ii)~We propose CIR-DDG, a 1\,537-parameter masked residual adapter that attaches to any \ddG{} predictor without retraining the backbone, and show that it improves all six evaluated backbones on antibody--antigen interface mutations while exactly conserving predictions elsewhere. (iii)~Through equal-capacity controls and feature-group ablations we attribute the gain to the information content of the descriptors rather than to the added parameters. (iv)~We demonstrate zero-shot transfer of the learned correction to an independent SARS-CoV-2 RBD--ACE2 deep-mutational-scan benchmark, clarifying that interface geometric sensitivity transfers whereas thermodynamic sign and scale do not.

\section{Materials and methods}

\subsection{Problem formulation}

For a wild-type complex with binding free energy $\Delta G_{\mathrm{wt}}$ and its mutant with $\Delta G_{\mathrm{mut}}$, we define
\begin{equation}
\ddG = \Delta G_{\mathrm{mut}}-\Delta G_{\mathrm{wt}}.
\label{eq:ddg}
\end{equation}
Negative values indicate affinity-enhancing (stabilizing) mutations under this convention, whereas positive values indicate affinity loss. For interpretation, the effect at site $i$ can be viewed as an approximate sum of a monomer-context contribution and a cross-chain interface contribution,
\begin{equation}
\ddG_i \approx \ddG_i^{\mathrm{mono}}+\ddG_i^{\mathrm{int}}.
\label{eq:decomp}
\end{equation}
The first term reflects changes in the local environment of the mutated chain; the second reflects geometric and chemical complementarity with the binding partner. The sum is approximate because the two contributions are not strictly separable; Equation~\eqref{eq:decomp} serves only to motivate the module design below, and no downstream quantity is computed from it.

This decomposition motivates three design requirements for a corrective module. First, the module should act only where the interface contribution is expected to be large, so that the behaviour of the base predictor elsewhere is inherited rather than relearned. Second, its inputs should be restricted to quantities that plausibly encode the interface term (distances, contacts and partner chemistry), so that limited supervision is spent on a low-dimensional, physically meaningful hypothesis space. Third, its capacity should be small enough that any observed gain can be attributed to the information content of the inputs rather than to additional trainable parameters, a requirement we verify directly with an equal-capacity control.

Let $\widehat{y}_{\mathrm{base}}$ be the out-of-fold prediction of a backbone and $\mathbf{g}\in\mathbb{R}^{22}$ the explicit interface descriptor defined below. On antibody--antigen interface samples, we fit ridge regressions ($\alpha=1$) within the complex-level five-fold protocol to estimate $R^2(\widehat{y}_{\mathrm{base}}\!\rightarrow y)$, $R^2(\mathbf{g}\!\rightarrow y)$ and $R^2([\widehat{y}_{\mathrm{base}},\mathbf{g}]\!\rightarrow y)$. The incremental contribution is
\begin{equation}
\Delta R^2 = R^2([\widehat{y}_{\mathrm{base}},\mathbf{g}]\!\rightarrow y)
- R^2(\widehat{y}_{\mathrm{base}}\!\rightarrow y).
\end{equation}
We also regress each geometric dimension on $\widehat{y}_{\mathrm{base}}$, truncate negative cross-validated $R^2$ values at zero and average across the 22 dimensions. This quantity, denoted $R^2(\mathrm{pred}\!\rightarrow\!\mathrm{geom})$, measures how much geometry is linearly decodable from the scalar prediction. Together, $\Delta R^2$ and $R^2(\mathrm{pred}\!\rightarrow\!\mathrm{geom})$ formalize output-level under-encoding.

\subsection{Dataset and evaluation}

All reported experiments use SKEMPI~2.0 \citep{jankauskaite2019skempi} measurements after the filtering in the supplied experimental pipeline: 4947 single-point and 1759 multiple-point entries from 343 complexes. A fixed complex-to-fold mapping assigns 69, 69, 69, 69 and 67 complexes to the five outer folds \citep{bushuiev2024learning, yu2025cath}, so mutations from the same complex never appear in both training and test partitions. Because splitting is performed on whole complexes, entries that share a protein--protein system (and therefore identical interface sequences) are always confined to the same partition, so no identical or near-identical interface can straddle training and test sets. The random seed is 2026.

Antibody-containing complexes were identified from SKEMPI complex annotations together with antibody-name registries, yielding an antibody-expanded cohort of 647 entries. Within these complexes, the principal target cohort collects single-point substitutions whose mutated residue lies within 5~\AA{} of the partner chain (minimum heavy-atom distance), irrespective of which side of the interface carries the substitution. The cohort contains 519 evaluable entries for five backbones and 517 for ESM-IF, for which two base predictions are unavailable; 303 entries are antibody-side (268 in CDRs and 35 in framework regions) and 216 are antigen-side. Multiple-point entries and non-interface entries are retained for training and for global evaluation but never activate the correction. Results are reported for all SKEMPI entries, single- and multiple-point subsets, and per-complex aggregates. Per-complex metrics are averaged only over complexes with at least ten measurements.

The six backbones are ProteinMPNN \citep{dauparas2022robust}, ESM-IF \citep{hsu2022learning}, RDE-Network \citep{luo2023rotamer}, DiffAffinity \citep{liu2023predicting}, DDAffinity \citep{yu2024ddaffinity} and Vanilla Pythia-PPI \citep{tao2025reliable}. ProteinMPNN and ESM-IF use zero-shot scores; the remaining backbones use five-fold trained weights fitted under the same complex-level split. We report Pearson correlation ($r$), Spearman rank correlation ($\rho$), RMSE, MAE and AUROC, treating $\ddG<0$ as the positive affinity-enhancing class. RMSE and MAE are computed after a univariate linear calibration between predicted and measured values.

\subsection{Cross-chain interface descriptors}

For a mutated residue $i$, let $A$ be the set of partner-chain heavy atoms and let $\mathrm{atoms}(i)$ be the heavy atoms of residue $i$, all taken from the wild-type complex. The minimum cross-chain distance is
\begin{equation}
d_{\min}(i)=\min_{a\in A,\,b\in\mathrm{atoms}(i)}\lVert\mathbf{x}_b-\mathbf{x}_a\rVert_2.
\label{eq:dmin}
\end{equation}
The descriptor $\mathbf{g}(i)\in\mathbb{R}^{22}$ comprises three groups, enumerated in full order in Supplementary Table~S3. The first group contains nine distance statistics: $d_{\min}$ itself, its log and reciprocal transforms and an exponential proximity kernel, the mean and standard deviation of the five nearest residue-level distances, and the median, first and third quartiles of the distance distribution over all partner residues. The second group contains nine contact features: log-transformed counts of partner residues within 4, 5, 6, 8 and 10~\AA{} of the site, the ratio of the 5~\AA{} to the 10~\AA{} residue counts, and log-transformed cross-chain heavy-atom pair counts within 4, 5 and 6~\AA{}; for a shell radius $r$,
\begin{equation}
c_r(i)=\left|\left\{(b,a):b\in\mathrm{atoms}(i),a\in A,
\lVert\mathbf{x}_b-\mathbf{x}_a\rVert_2\le r\right\}\right|.
\end{equation}
The third group contains four site--partner context terms: the log-transformed number of heavy atoms at the site, the log-transformed number of partner-chain residues, and the hydrophobic and charged fractions of partner residues within 5~\AA{} of the site, using the classes hydrophobic $\{$A,I,L,M,F,W,V,Y$\}$ and charged $\{$D,E,K,R,H$\}$. Each dimension is a distance, count, density or chemical frequency rather than a learned embedding. The choice follows long-standing evidence \citep{kortemme2002simple, pires2014mcsm, rodrigues2019mcsm} that interfacial hot spots, atomic-distance patterns and cross-partner contact networks are informative for mutation-induced affinity changes. Descriptors are computed once per site from the wild-type structure and standardized with the mean and standard deviation of the training complexes within each fold.

\subsection{Cross-chain interaction residual}

Given a base prediction, CIR-DDG produces
\begin{equation}
\widehat{y}(x)=\widehat{y}_{\mathrm{base}}(x)
+s(x)m(x)r_{\phi}(\mathbf{g}(x)),
\label{eq:cir}
\end{equation}
where $r_{\phi}$ is a $22\!\rightarrow\!64\!\rightarrow\!1$ multilayer perceptron with a GELU nonlinearity and dropout 0.1, totalling 1\,537 parameters, several orders of magnitude fewer than any evaluated backbone. Writing $\mathcal{S}_{\mathrm{ab}}$ for the set of single-point mutations in antibody--antigen complexes, the binary mask is
\begin{equation}
m(x)=
\begin{cases}
1, & x\in\mathcal{S}_{\mathrm{ab}}\ \text{and}\ d_{\min}(x)\le5\ \text{\AA},\\
0, & \text{otherwise},
\end{cases}
\end{equation}
so predictions outside the target cohort are exactly preserved. The sign $s(x)\in\{+1,-1\}$ makes the correction antisymmetric for datasets containing forward and reverse mutations; all filtered SKEMPI~2.0 entries are forward (wild-type to mutant), so $s=+1$ throughout this study.

\begin{figure*}[!tb]
\centering
\includegraphics[width=\textwidth]{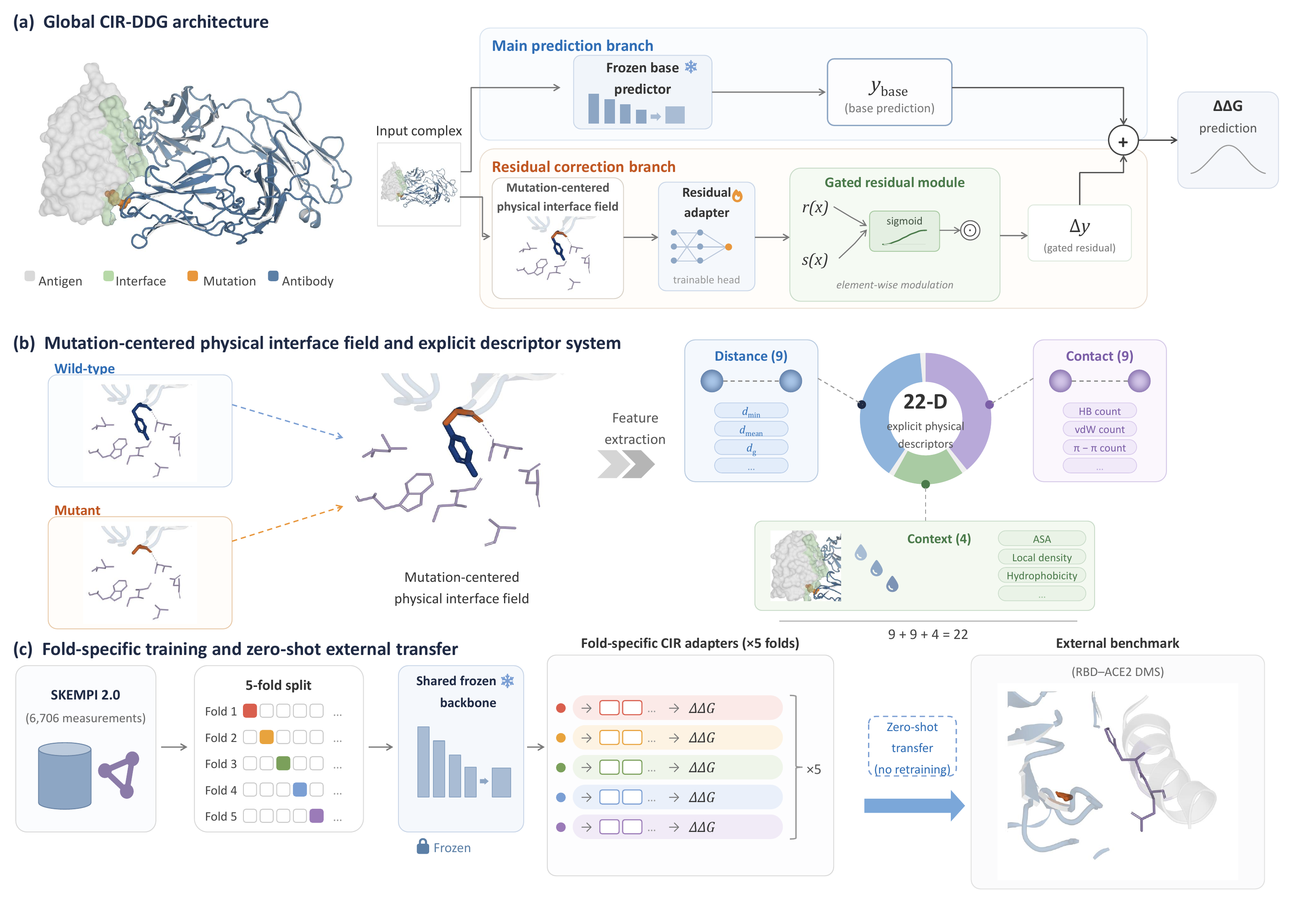}
\caption{Overview of CIR-DDG. (a)~A frozen base predictor supplies the main prediction; a mutation-centred physical interface field feeds a lightweight gated residual adapter that corrects it. (b)~Wild-type and mutant states at the mutation site are summarised by 22 explicit physical descriptors (nine distance, nine contact, four context). (c)~Fold-specific adapters trained on a complex-level five-fold split of SKEMPI~2.0 transfer zero-shot to the external RBD--ACE2 DMS benchmark.\label{fig:overview}}
\alttext{Three-panel flow diagram. Panel a: an antibody--antigen complex feeds a frozen base-predictor branch and a residual-correction branch containing a mutation-centred physical interface field, a lightweight adapter and a gated residual module; the two branches merge at a summation node to give the final delta-delta-G prediction. Panel b: wild-type and mutant states at the mutation site feed the extraction of 22 explicit physical descriptors grouped as nine distance, nine contact and four context features. Panel c: a five-fold split of SKEMPI trains fold-specific adapters on a shared frozen backbone, which transfers zero-shot to an external RBD--ACE2 benchmark.}
\end{figure*}

Figure~\ref{fig:overview} summarizes the method. CIR-DDG is intentionally additive: it represents a compact correction to a heterogeneous base predictor rather than a replacement encoder.

\subsection{Training objective and optimization}

For training set $D$ and its activated interface subset $D_{\mathrm{int}}$, the objective is
\begin{equation}
\mathcal{L}=\mathcal{L}_{\mathrm{MSE}}+\lambda_{\mathrm{int}}\mathcal{L}_{\mathrm{int}}
+\lambda_{\mathrm{con}}\mathcal{L}_{\mathrm{consist}},
\end{equation}
with
\begin{align}
\mathcal{L}_{\mathrm{MSE}}&=\frac{1}{|D|}\sum_{x\in D}(\widehat{y}(x)-y(x))^2,\\
\mathcal{L}_{\mathrm{int}}&=\frac{1}{|D_{\mathrm{int}}|}\sum_{x\in D_{\mathrm{int}}}(\widehat{y}(x)-y(x))^2,\\
\mathcal{L}_{\mathrm{consist}}&=1-r_{\mathrm{Pearson}}(\widehat{y},y).
\end{align}
We set $\lambda_{\mathrm{int}}=1.0$ and $\lambda_{\mathrm{con}}=0.2$. Because the residual is activated for only approximately 8\% of the full dataset, the consistency term discourages a local ranking gain that disrupts the global response scale.

Within each outer fold, training complexes are divided into three inner folds and one inner fold is reserved for checkpoint selection. The checkpoint-selection score is
\begin{equation}
V_{\mathrm{val}}=\rho_{\mathrm{int}}-\beta\max(0,r_{\mathrm{base}}-r_{\mathrm{CIR}}),
\end{equation}
where $\rho_{\mathrm{int}}$ is the Spearman correlation on the interface subset of the reserved inner fold, the two $r$ terms are global Pearson correlations on the same inner fold, and $\beta=3.0$ weights the penalty for global-response degradation. Training uses AdamW for at most 150 epochs, patience 30, learning rate $5\times10^{-4}$, weight decay $10^{-4}$ and gradient-norm clipping at 1.0. After out-of-fold prediction, a constant offset recentres the corrected interface predictions on the mean of the base predictions of the same cohort, so that the adapter introduces no constant shift between base and corrected outputs. For inverse-folding backbones, the pre-trained encoder remains frozen while the prediction and residual heads are fitted; trainable end-to-end backbones are fine-tuned jointly according to the supplied pipeline.

\section{Results}

\subsection{Antibody--antigen interface performance}

The activated antibody--antigen interface cohort is the primary test of CIR-DDG. Across all six backbones, every reported metric changes in the favourable direction (Table~\ref{tab:interface}, Supplementary Fig.~S4). Spearman $\rho$ increases by 0.0346--0.1296, Pearson $r$ by 0.0150--0.1011, RMSE decreases by 0.0074--0.0408~\kcalmol, MAE decreases by 0.0062--0.0481~\kcalmol{} and AUROC increases by 0.0113--0.0572. The largest rank-correlation gain occurs for ProteinMPNN (0.1716 to 0.3012), followed by RDE-Network (0.3621 to 0.4644).

The two zero-shot inverse-folding backbones, whose scores derive from structure-conditioned sequence likelihoods rather than from affinity supervision, show the largest relative improvements: ProteinMPNN and ESM-IF gain 0.1296 and 0.0671 in Spearman $\rho$, respectively. The four supervised or side-chain-aware backbones, which already encode parts of the interface signal, gain less in absolute terms but still improve on every metric, and the strongest base model (Vanilla Pythia-PPI, $\rho=0.4699$) retains its lead after correction ($\rho=0.5045$). CIR-DDG therefore does not merely rescue weak backbones: it improves every backbone without altering their relative ranking; the consistency of this gain across all complementary analyses is summarised in Supplementary Fig.~S2.

\begin{table*}[!tb]
\caption{Performance on antibody--antigen interface mutations. The cohort contains only single-point mutations. Higher is better for $r$, $\rho$ and AUROC; lower is better for RMSE and MAE.\label{tab:interface}}
\centering\footnotesize
\tabcolsep=5pt
\begin{tabular*}{\textwidth}{@{\extracolsep{\fill}}lccccc@{\extracolsep{\fill}}}
\toprule
Model & Pearson $r$ & Spearman $\rho$ & RMSE & MAE & AUROC\\
\midrule
ProteinMPNN & 0.1809 & 0.1716 & 1.6173 & 1.2061 & 0.6389\\
ProteinMPNN + CIR-DDG & \textbf{0.2819} & \textbf{0.3012} & \textbf{1.5778} & \textbf{1.1733} & \textbf{0.6961}\\
ESM-IF & 0.0483 & 0.1187 & 1.6447 & 1.2123 & 0.6398\\
ESM-IF + CIR-DDG & \textbf{0.1060} & \textbf{0.1858} & \textbf{1.6373} & \textbf{1.2061} & \textbf{0.6622}\\
RDE-Network & 0.5215 & 0.3621 & 1.4031 & 1.0850 & 0.6887\\
RDE-Network + CIR-DDG & \textbf{0.5601} & \textbf{0.4644} & \textbf{1.3623} & \textbf{1.0495} & \textbf{0.7230}\\
DiffAffinity & 0.5545 & 0.4389 & 1.3685 & 1.0457 & 0.7376\\
DiffAffinity + CIR-DDG & \textbf{0.5868} & \textbf{0.5098} & \textbf{1.3315} & \textbf{0.9976} & \textbf{0.7611}\\
DDAffinity & 0.5015 & 0.3451 & 1.4227 & 1.0930 & 0.6712\\
DDAffinity + CIR-DDG & \textbf{0.5277} & \textbf{0.3885} & \textbf{1.3968} & \textbf{1.0687} & \textbf{0.6906}\\
Vanilla Pythia-PPI & 0.5946 & 0.4699 & 1.3222 & 0.9964 & 0.7525\\
Vanilla Pythia-PPI + CIR-DDG & \textbf{0.6096} & \textbf{0.5045} & \textbf{1.3036} & \textbf{0.9755} & \textbf{0.7639}\\
\botrule
\end{tabular*}
\begin{tablenotes}
\item RMSE and MAE are in \kcalmol{} and follow the supplied linear-calibration procedure.
\end{tablenotes}
\end{table*}

\subsection{Scope conservation}

Pooled over each backbone's full evaluable entry set, CIR-DDG increases Spearman $\rho$ by 0.0018--0.0112; on single-point mutations, the increase is 0.0026--0.0159. Multiple-point metrics are unchanged because the residual mask is inactive. The corresponding all-, single- and multiple-point results are provided in Supplementary Table~S1. When metrics are first computed per complex and then averaged, changes are smaller and some error metrics fluctuate by approximately $10^{-3}$ (Supplementary Table~S2). These results are consistent with a local correction acting on roughly 8\% of measurements: gains are diluted when averaged over the many unmodified samples, while the pre-existing behaviour of the backbone is retained outside the specified interface cohort.

\subsection{Probing analysis}

The cross-validated ridge analysis tests whether the interface descriptors remain informative after conditioning on the scalar base prediction (Table~\ref{tab:probe}, Supplementary Fig.~S3). Geometry alone explains a consistent fraction of variance ($R^2=0.1167$--$0.1175$), nearly identical across backbones because the descriptor is model-agnostic and differs only through each backbone's evaluable entry set. Adding geometry to the base prediction yields a positive $\Delta R^2$ for every backbone, ranging from 0.0205 for DiffAffinity to 0.1533 for ESM-IF. In contrast, the mean geometry decodable from the base scalar is only 0--0.0478. The inverse-folding backbones combine the largest $\Delta R^2$ with undecodable geometry ($R^2(\mathrm{pred}\!\rightarrow\!\mathrm{geom})=0$), mirroring their large empirical gains, whereas supervised backbones occupy an intermediate regime in which a small part of the descriptor is linearly present in the output yet a positive increment remains. The result supports output-level complementarity between explicit geometry and the base score.

\begin{table*}[!tb]
\caption{Complex-level five-fold ridge analysis on antibody--antigen interface mutations. Negative $R^2$ indicates performance below a mean predictor.\label{tab:probe}}
\centering\footnotesize
\begin{tabular*}{\textwidth}{@{\extracolsep{\fill}}lrrrrrr@{\extracolsep{\fill}}}
\toprule
Model & $N$ & $R^2$(base) & $R^2$(geom) & $R^2$(both) & $\Delta R^2$ & $R^2$(pred$\rightarrow$geom)\\
\midrule
ProteinMPNN & 519 & $-0.0209$ & 0.1175 & 0.1219 & 0.1428 & 0.0000\\
ESM-IF & 517 & $-0.0478$ & 0.1167 & 0.1055 & 0.1533 & 0.0000\\
RDE-Network & 519 & 0.2443 & 0.1175 & 0.2747 & 0.0304 & 0.0345\\
DiffAffinity & 519 & 0.2833 & 0.1175 & 0.3038 & 0.0205 & 0.0478\\
DDAffinity & 519 & 0.2432 & 0.1175 & 0.2849 & 0.0417 & 0.0276\\
Vanilla Pythia-PPI & 519 & 0.3443 & 0.1175 & 0.3670 & 0.0227 & 0.0347\\
\botrule
\end{tabular*}
\end{table*}

\subsection{Ablation study}

Replacing all 22 inputs with zeros controls for the parameter count and optimization path of the residual head. This control changes interface Spearman $\rho$ by only $-0.0021$ to $+0.0043$ relative to the corresponding base model (Table~\ref{tab:ablation}, Supplementary Fig.~S1), whereas the full descriptor improves all six backbones by 0.0365--0.1117 in this ablation run. The gain is therefore attributable to the information content of the descriptors rather than to the addition of a small trainable head. Among partial descriptors, contact plus chemistry is strongest for five backbones, while chemistry alone is strongest for ESM-IF. The full descriptor is best for every backbone. Distance features alone recover 44--62\% of the full gain for ProteinMPNN, RDE-Network and DiffAffinity, whereas the gain is carried mainly by chemistry for ESM-IF and by contact features for DDAffinity and Vanilla Pythia-PPI, indicating that distance, contact abundance and partner chemistry carry complementary rather than redundant signal.

\begin{table*}[!tb]
\caption{Interface Spearman $\rho$ under equal-capacity and feature ablations.\label{tab:ablation}}
\centering\scriptsize
\tabcolsep=3.5pt
\begin{tabular*}{\textwidth}{@{\extracolsep{\fill}}lrrrrrrr@{\extracolsep{\fill}}}
\toprule
Model & Base & Full 22D & Zero 22D & Distance 9D & Contact 9D & Chemistry 4D & Contact+chem. 13D\\
\midrule
ProteinMPNN & 0.1716 & \textbf{0.2637} & 0.1722 & 0.2121 & 0.2224 & 0.2170 & 0.2558\\
ESM-IF & 0.1187 & \textbf{0.1552} & 0.1202 & 0.1324 & 0.1093 & 0.1481 & 0.1221\\
RDE-Network & 0.3621 & \textbf{0.4738} & 0.3601 & 0.4314 & 0.4450 & 0.3757 & 0.4528\\
DiffAffinity & 0.4389 & \textbf{0.5153} & 0.4371 & 0.4864 & 0.4879 & 0.4532 & 0.5101\\
DDAffinity & 0.3451 & \textbf{0.4210} & 0.3494 & 0.3649 & 0.4041 & 0.3141 & 0.4050\\
Vanilla Pythia-PPI & 0.4699 & \textbf{0.5388} & 0.4693 & 0.4667 & 0.5207 & 0.4555 & 0.5289\\
\botrule
\end{tabular*}
\end{table*}

\subsection{External validation}

\begin{figure*}[!tb]
\centering
\includegraphics[width=\textwidth]{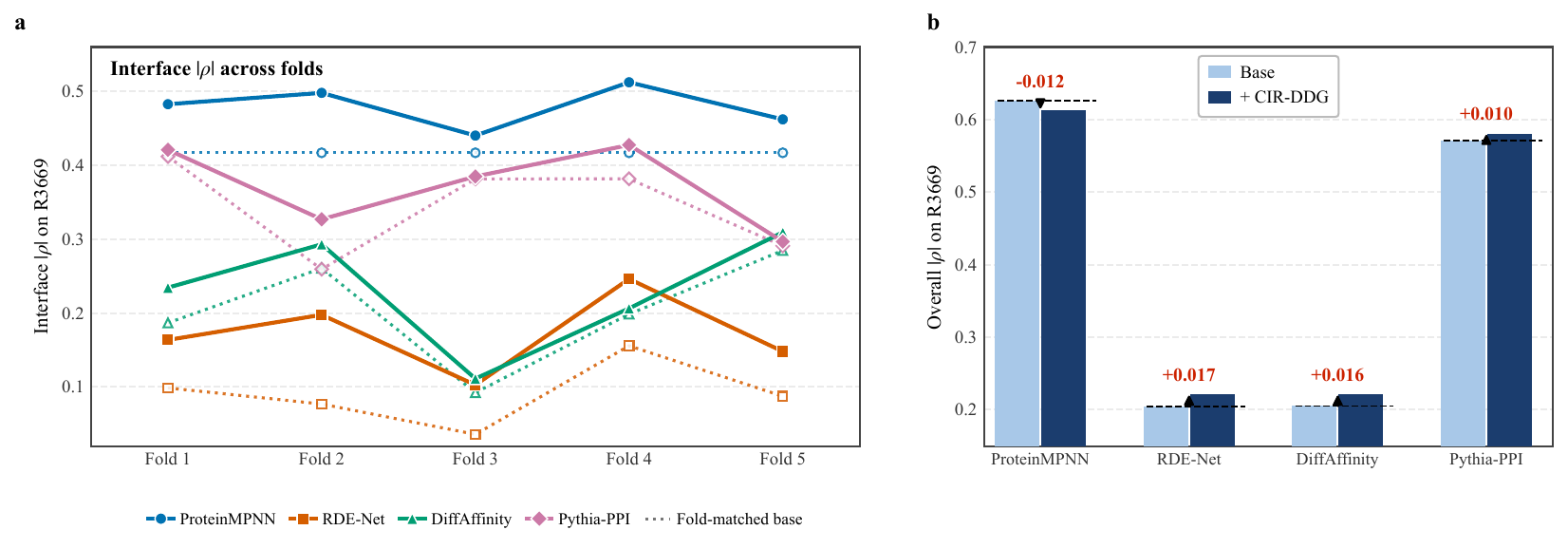}
\caption{Zero-shot transfer to the SARS-CoV-2 RBD--ACE2 deep-mutational-scan benchmark. (a)~Interface $|\rho|$ of the five fold-specific models (solid lines) against fold-matched base values (dotted lines) for each backbone. (b)~Pooled overall $|\rho|$ before (light blue) and after (dark blue) correction; red arrows mark the per-backbone change. Absolute correlations are shown because the DMS score is opposite in sign to \ddG{}. Values are listed in Table~\ref{tab:external} and Supplementary Table~S4.\label{fig:external}}
\alttext{Left, line chart of interface absolute Spearman correlation across the five folds for four backbones, solid lines showing corrected models and dotted lines showing fold-matched base values; every corrected point lies above its fold-matched base value. Right, paired bars of pooled overall absolute correlation before and after correction, increasing for three backbones and slightly decreasing for ProteinMPNN.}
\end{figure*}

The external benchmark asks whether the geometric correction learned on SKEMPI antibody--antigen interfaces transfers to an interface that shares neither complex, fold, measurement type nor organism with the training data. We assembled it from the yeast-display deep mutational scanning study of \citet{starr2022shifting}, which quantified ACE2-binding effects for near-saturating single substitutions of the ancestral SARS-CoV-2 spike receptor-binding domain (RBD); the processed cohort, denoted R3669, contains 3669 single-point substitutions at 194 RBD positions, with the crystal structure PDB 6M0J \citep{lan2020structure} providing the wild-type geometry (chain E mutated, chain A as partner). 6M0J post-dates SKEMPI~2.0 and shares no complex with the training data, and because the SARS-CoV-2 RBD itself post-dates SKEMPI~2.0, no sequence variant of this interface can occur in training. For each of the five complex-level folds, the paired base checkpoint and fold-specific CIR-DDG adapter were applied without retraining, recalibration or access to R3669 labels, and substitutions at the 21 positions with $d_{\min}\le5$~\AA{} relative to ACE2 form the external interface cohort (protocol details and the position list are given with Supplementary Table~S4). Because the DMS binding score is opposite in sign to \ddG{} (Eq.~\eqref{eq:ddg}), all correlations are negative and absolute Spearman values $|\rho|$ are compared; a larger $|\rho|$ indicates better recovery of the experimental ranking. Table~\ref{tab:external} and Fig.~\ref{fig:external} report the results.

At the interface (the cohort on which the correction acts), $|\rho|$ increased for all four evaluable backbones, by 0.0263--0.0807. The largest absolute gain occurred for RDE-Network ($-0.0909$ to $-0.1716$), whose base ranking at the viral interface is close to random, followed by ProteinMPNN ($-0.4169$ to $-0.4790$). The improvement is consistent at fold granularity: for every backbone, each of the five fold-specific adapters individually increased interface $|\rho|$ relative to its paired base model (20 of 20 comparisons; Supplementary Table~S4). Because no parameter was fitted on R3669, these gains measure zero-shot transfer of the learned geometry-to-correction mapping to an interface type absent from training.

Pooled over all 3669 substitutions, $|\rho|$ increased for RDE-Network, DiffAffinity and Vanilla Pythia-PPI and decreased slightly for ProteinMPNN (0.6256 to 0.6136). The ProteinMPNN pattern is a consequence of the design rather than of inconsistent behaviour: its strong pooled anti-correlation is dominated by non-interface substitutions, which the mask leaves unchanged, whereas the correction improves ranking specifically within the interface cohort. The pooled DMS ranking additionally conflates binding with RBD expression and folding constraints at buried sites, which lie outside the scope of an interface-targeted correction. External validation therefore supports the narrower claim that explicit cross-chain geometry transfers across interface types, while also showing that thermodynamic calibration and sign do not transfer automatically: both would need to be re-established before prospective ranking use on DMS-scale data. The five analyses are mutually consistent: the gain concentrates on the interface-targeted quantities and remains near zero where the correction is designed to be inactive (Supplementary Fig.~S2).

\begin{table*}[!tb]
\caption{External validation on 3669 SARS-CoV-2 RBD--ACE2 deep-mutational-scan substitutions (6M0J). Spearman $\rho$ against the DMS binding score (opposite in sign to \ddG{}; larger $|\rho|$ indicates better ranking), averaged over the five fold-specific models.\label{tab:external}}
\centering\footnotesize
\tabcolsep=5pt
\begin{tabular*}{\textwidth}{@{\extracolsep{\fill}}lccccc@{\extracolsep{\fill}}}
\toprule
Model & Overall $\rho$ (base) & Overall $\rho$ (+CIR-DDG) & Interface $\rho$ (base) & Interface $\rho$ (+CIR-DDG) & $\Delta|\rho|$ interface\\
\midrule
ProteinMPNN & $-0.6256$ & $-0.6136$ & $-0.4169$ & \textbf{$-0.4790$} & $+0.0621$\\
RDE-Network & $-0.2040$ & $-0.2212$ & $-0.0909$ & \textbf{$-0.1716$} & $+0.0807$\\
DiffAffinity & $-0.2051$ & $-0.2211$ & $-0.2045$ & \textbf{$-0.2308$} & $+0.0263$\\
Vanilla Pythia-PPI & $-0.5705$ & $-0.5807$ & $-0.3450$ & \textbf{$-0.3713$} & $+0.0263$\\
\botrule
\end{tabular*}
\begin{tablenotes}
\item ProteinMPNN is zero-shot; its base prediction is shared across folds and only the adapter varies. ESM-IF and DDAffinity could not be evaluated on this benchmark (Supplementary protocol note). Per-fold values are given in Supplementary Table~S4.
\end{tablenotes}
\end{table*}

\subsection{Case study}

We examine HyHEL-10 Fab bound to hen egg-white lysozyme (3HFM, chains H/L/Y), which belongs to outer fold~2 and contains 96 interface measurements with geometry. For the fold-specific RDE-Network checkpoint, CIR-DDG increases Spearman $\rho$ from 0.2731 to 0.3741 and reduces MAE from 1.5633 to 1.3252~\kcalmol. Across the measurements, the correction has the same sign as the base residual $y-\widehat{y}_{\mathrm{base}}$ in 51.85\% of cases.

\begin{figure}[!tb]
\centering
\begingroup
\setlength{\fboxsep}{0pt}\setlength{\fboxrule}{0.4pt}%
\begin{minipage}[t]{0.49\columnwidth}
\centering\fcolorbox{black!20}{white}{\includegraphics[width=0.985\linewidth]{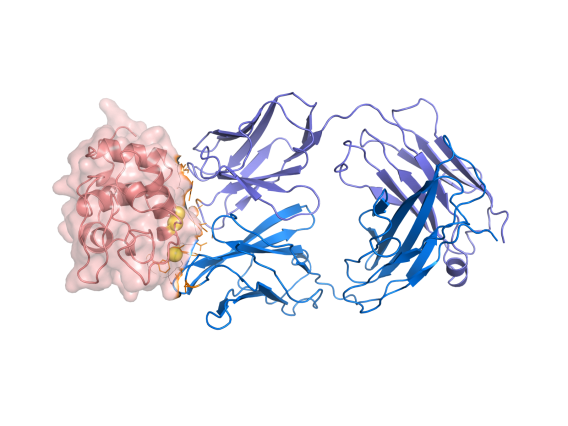}}\\[1pt]
{\footnotesize\textbf{(a)} Overall complex}
\end{minipage}\hfill
\begin{minipage}[t]{0.49\columnwidth}
\centering\fcolorbox{black!20}{white}{\includegraphics[width=0.985\linewidth]{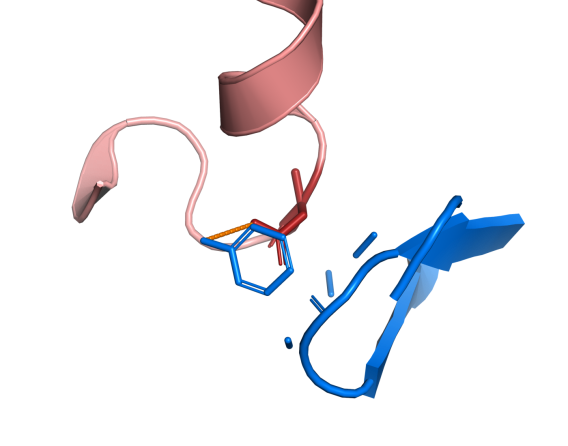}}\\[1pt]
{\footnotesize\textbf{(b)} D101(Y): correction 0.990, $d_{\min}=2.311$~\AA}
\end{minipage}

\vspace{3pt}
\begin{minipage}[t]{0.49\columnwidth}
\centering\fcolorbox{black!20}{white}{\includegraphics[width=0.985\linewidth]{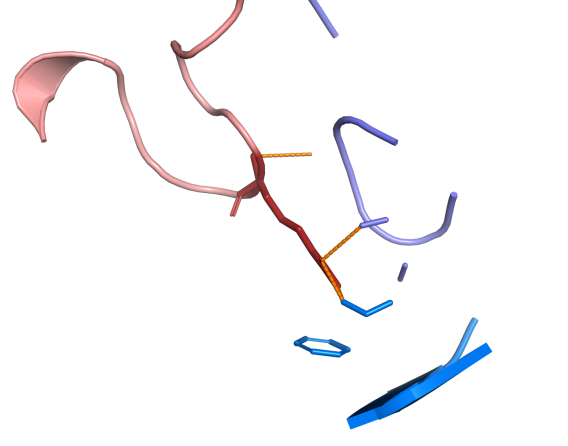}}\\[1pt]
{\footnotesize\textbf{(c)} R21(Y): correction 0.882, $d_{\min}=2.512$~\AA}
\end{minipage}\hfill
\begin{minipage}[t]{0.49\columnwidth}
\centering\fcolorbox{black!20}{white}{\includegraphics[width=0.985\linewidth]{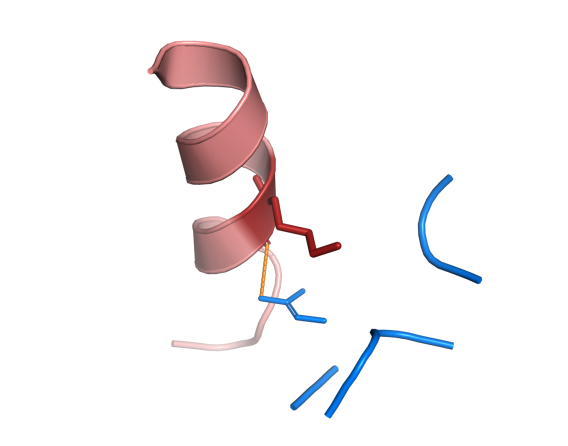}}\\[1pt]
{\footnotesize\textbf{(d)} K97(Y): correction 0.630, $d_{\min}=3.001$~\AA}
\end{minipage}
\endgroup
\caption{Cross-chain geometry of representative sites in the 3HFM HyHEL-10--lysozyme complex. Antibody chains are blue, the antigen is pink, target residues are dark red and selected cross-chain contacts are orange dashed lines.\label{fig:3hfm}}
\alttext{Four molecular renderings show the full HyHEL-10 antibody bound to lysozyme and close views of lysozyme sites D101, R21 and K97 near antibody atoms. Orange dashed lines mark short cross-chain contacts at each highlighted site.}
\end{figure}

Three sites on the lysozyme (antigen) chain illustrate the relationship between correction and local contact geometry (Table~\ref{tab:3hfm}; Fig.~\ref{fig:3hfm}). Mean corrections at D101(Y), R21(Y) and K97(Y) are positive, as are their mean base residuals, indicating that RDE-Network underestimates destabilization at these positions on average. D101(Y), the lysozyme residue closest to the antibody with $d_{\min}=2.31$~\AA, is supported by 19 measurements rather than by a single selected substitution, and its correction of nearly 1~\kcalmol{} is consistent with a dense cross-chain contact shell that the base score does not fully price.

\begin{table}[!tb]
\caption{Aggregated statistics for representative sites in 3HFM. The base residual is $y-\widehat{y}_{\mathrm{base}}$.\label{tab:3hfm}}
\centering\scriptsize
\begin{tabular*}{\columnwidth}{@{\extracolsep{\fill}}lrrrr@{\extracolsep{\fill}}}
\toprule
Site & $N$ & $d_{\min}$ (\AA) & Mean corr. & Mean residual\\
\midrule
D101(Y) & 19 & 2.3114 & 0.9901 & 1.0114\\
R21(Y) & 18 & 2.5118 & 0.8819 & 1.4065\\
K97(Y) & 8 & 3.0010 & 0.6296 & 2.8194\\
\botrule
\end{tabular*}
\end{table}

\section{Conclusion}

CIR-DDG treats \ddG{} prediction as output-level repair: a frozen base predictor is combined with a masked additive residual driven by 22 explicit descriptors of cross-chain distance, contact density and partner chemistry. On SKEMPI~2.0 antibody--antigen interface mutations, the same lightweight adapter improved all six evaluated backbones on every reported metric, and cross-validated probing, the equal-capacity control and feature ablations attribute the gain to the information content of the descriptors rather than to the added parameters. Because the scope mask restricts the correction to its intended cohort, predictions elsewhere are conserved by design. Since the module reads only a scalar base prediction and a fixed descriptor, it can be attached to predictors whose internals are unavailable, retrained in minutes and audited feature by feature.

The learned geometry-to-correction mapping also transferred zero-shot to an independent SARS-CoV-2 RBD--ACE2 deep-mutational-scan benchmark, improving interface ranking for all four evaluable backbones, although absolute calibration, sign and pooled-metric behaviour did not transfer automatically; prospective use on a new interface family should therefore pair CIR-DDG with a small local calibration set. The method remains limited to single substitutions at protein--protein interfaces, relies on a single static complex structure without conformational ensembles or explicit solvent, and has so far been validated externally on one viral interface; extending the descriptor to ensemble geometry and multiple mutants, and broadening both training and external cohorts beyond antibody--antigen interfaces, are natural next steps. CIR-DDG thus offers a practical and interpretable route to interface-aware refinement of heterogeneous \ddG{} predictors under limited experimental supervision.

\section*{Acknowledgements}
During the preparation of this work, the authors used Kimi K3 (Moonshot AI) to check the manuscript text for grammatical errors. The authors reviewed and edited the content and take full responsibility for the integrity of this publication.

\section*{Supplementary data}
Supplementary data accompany this article.

\section*{Conflict of interest}
None declared.

\section*{Funding}
This work was supported by the Shanghai Special Program for Promoting High-Quality Industrial Development [250401]; the AI for Science Program of the Shanghai Municipal Commission of Economy and Informatization [2025-GZL-RGZN-BTBX-02013]; and the ECNU Multifunctional Platform for Innovation [001].

\section*{Data availability}
SKEMPI~2.0 is publicly available \citep{jankauskaite2019skempi,skempi2dataset}. The external benchmark derives from public deep-mutational-scanning measurements \citep{starr2022shifting,starr2022dmsdata} and PDB entry 6M0J \citep{lan2020structure,pdb6m0j}. The CIR-DDG source code, processed descriptors, fold assignments, released predictions and all experiment scripts are available at \url{https://github.com/ecnuabmlab/CIR-ddG}.

\bibliographystyle{oup-abbrvnat}
\bibliography{references}

@article{jankauskaite2019skempi,
  title={SKEMPI 2.0: an updated benchmark of changes in protein--protein binding energy, kinetics and thermodynamics upon mutation},
  author={Jankauskait{\.e}, Justina and Jim{\'e}nez-Garc{\'\i}a, Brian and Dapk{\=u}nas, Justas and Fern{\'a}ndez-Recio, Juan and Moal, Iain H},
  journal={Bioinformatics},
  volume={35},
  number={3},
  pages={462--469},
  year={2019},
  publisher={Oxford University Press}
}

@article{barlow2018flex,
  title={Flex ddG: Rosetta ensemble-based estimation of changes in protein--protein binding affinity upon mutation},
  author = {Barlow, Kyle A. and {{\'O} Conch{\'u}ir}, Shane and Thompson, Samuel and Suresh, Pooja and Lucas, James E. and Heinonen, Markus and Kortemme, Tanja},
  journal={The journal of physical chemistry B},
  volume={122},
  number={21},
  pages={5389--5399},
  year={2018},
  publisher={ACS Publications}
}

@inproceedings{hsu2022learning,
  title={Learning inverse folding from millions of predicted structures},
  author={Hsu, Chloe and Verkuil, Robert and Liu, Jason and Lin, Zeming and Hie, Brian and Sercu, Tom and Lerer, Adam and Rives, Alexander},
  booktitle={International conference on machine learning},
  pages={8946--8970},
  year={2022},
  organization={PMLR}
}

@article{dauparas2022robust,
  title={Robust deep learning--based protein sequence design using ProteinMPNN},
  author={Dauparas, Justas and Anishchenko, Ivan and Bennett, Nathaniel and Bai, Hua and Ragotte, Robert J and Milles, Lukas F and Wicky, Basile IM and Courbet, Alexis and de Haas, Rob J and Bethel, Neville and others},
  journal={Science},
  volume={378},
  number={6615},
  pages={49--56},
  year={2022},
  publisher={American Association for the Advancement of Science}
}

@inproceedings{luo2023rotamer,
  title={Rotamer Density Estimator is an Unsupervised Learner of the Effect of Mutations on Protein-Protein Interaction},
  author={Shitong Luo and Yufeng Su and Zuofan Wu and Chenpeng Su and Jian Peng and Jianzhu Ma},
  booktitle={The Eleventh International Conference on Learning Representations },
  year={2023}
}

@article{liu2023predicting,
  title={Predicting mutational effects on protein-protein binding via a side-chain diffusion probabilistic model},
  author={Liu, Shiwei and Zhu, Tian and Ren, Milong and Yu, Chungong and Bu, Dongbo and Zhang, Haicang},
  journal={Advances in Neural Information Processing Systems},
  volume={36},
  pages={48994--49005},
  year={2023}
}

@article{yu2024ddaffinity,
  title={DDAffinity: predicting the changes in binding affinity of multiple point mutations using protein 3D structure},
  author={Yu, Guanglei and Zhao, Qichang and Bi, Xuehua and Wang, Jianxin},
  journal={Bioinformatics},
  volume={40},
  number={Supplement\_1},
  pages={i418--i427},
  year={2024},
  publisher={Oxford University Press}
}

@inproceedings{bushuiev2024learning,
  title={Learning to design protein-protein interactions with enhanced generalization},
  author={Bushuiev, Anton and Bushuiev, Roman and Kouba, Petr and Filkin, Anatolii and Gabrielova, Marketa and Gabriel, Michal and Sedlar, Jiri and Pluskal, Tomas and Damborsky, Jiri and Mazurenko, Stanislav and others},
  booktitle={International Conference on Learning Representations},
  volume={2024},
  pages={21010--21035},
  year={2024}
}

@article{tao2025reliable,
  title={Reliable prediction of protein--protein binding affinity changes upon mutations with Pythia-PPI},
  author={Tao, Fangting and Sun, Jinyuan and Gao, Pengyue and Gao, George Fu and Wu, Bian},
  journal={National Science Review},
  volume={12},
  number={6},
  pages={nwaf231},
  year={2025},
  publisher={Oxford University Press}
}

@article{hummer2025investigating,
  title={Investigating the volume and diversity of data needed for generalizable antibody--antigen $\Delta$$\Delta$ G prediction},
  author={Hummer, Alissa M and Schneider, Constantin and Chinery, Lewis and Deane, Charlotte M},
  journal={Nature Computational Science},
  volume={5},
  number={8},
  pages={635--647},
  year={2025},
  publisher={Nature Publishing Group US New York}
}

@article{tsishyn2024quantification,
  title={Quantification of biases in predictions of protein--protein binding affinity changes upon mutations},
  author={Tsishyn, Matsvei and Pucci, Fabrizio and Rooman, Marianne},
  journal={Briefings in bioinformatics},
  volume={25},
  number={1},
  pages={bbad491},
  year={2024},
  publisher={Oxford University Press}
}

@article{kortemme2002simple,
  title={A simple physical model for binding energy hot spots in protein--protein complexes},
  author={Kortemme, Tanja and Baker, David},
  journal={Proceedings of the National Academy of Sciences},
  volume={99},
  number={22},
  pages={14116--14121},
  year={2002},
  publisher={National Academy of Sciences}
}

@article{delgado2019foldx,
  title={FoldX 5.0: working with RNA, small molecules and a new graphical interface},
  author={Delgado, Javier and Radusky, Leandro G and Cianferoni, Damiano and Serrano, Luis},
  journal={Bioinformatics},
  volume={35},
  number={20},
  pages={4168--4169},
  year={2019},
  publisher={Oxford University Press}
}

@article{dehouck2013beatmusic,
  title={BeAtMuSiC: prediction of changes in protein--protein binding affinity on mutations},
  author={Dehouck, Yves and Kwasigroch, Jean Marc and Rooman, Marianne and Gilis, Dimitri},
  journal={Nucleic acids research},
  volume={41},
  number={W1},
  pages={W333--W339},
  year={2013},
  publisher={Oxford University Press}
}

@article{pires2014mcsm,
  title={mCSM: predicting the effects of mutations in proteins using graph-based signatures},
  author={Pires, Douglas EV and Ascher, David B and Blundell, Tom L},
  journal={Bioinformatics},
  volume={30},
  number={3},
  pages={335--342},
  year={2014},
  publisher={Oxford University Press}
}

@article{rodrigues2019mcsm,
  title={mCSM-PPI2: predicting the effects of mutations on protein--protein interactions},
  author={Rodrigues, Carlos HM and Myung, Yoochan and Pires, Douglas EV and Ascher, David B},
  journal={Nucleic acids research},
  volume={47},
  number={W1},
  pages={W338--W344},
  year={2019},
  publisher={Oxford University Press}
}

@inproceedings{wu2024learning,
  title={Learning to Predict Mutational Effects of Protein-Protein Interactions by Microenvironment-aware Hierarchical Prompt Learning},
  author={Lirong Wu and Yijun Tian and Haitao Lin and Yufei Huang and Siyuan Li and Nitesh V Chawla and Stan Z. Li},
  booktitle={Forty-first International Conference on Machine Learning},
  year={2024}
}

@article{yu2025cath,
  title={CATH-ddG: towards robust mutation effect prediction on protein--protein interactions out of CATH homologous superfamily},
  author={Yu, Guanglei and Bi, Xuehua and Ma, Teng and Li, Yaohang and Wang, Jianxin},
  journal={Bioinformatics},
  volume={41},
  number={Supplement\_1},
  pages={i362--i372},
  year={2025},
  publisher={Oxford University Press}
}

@article{makowski2022co,
  title={Co-optimization of therapeutic antibody affinity and specificity using machine learning models that generalize to novel mutational space},
  author={Makowski, Emily K and Kinnunen, Patrick C and Huang, Jie and Wu, Lina and Smith, Matthew D and Wang, Tiexin and Desai, Alec A and Streu, Craig N and Zhang, Yulei and Zupancic, Jennifer M and others},
  journal={Nature communications},
  volume={13},
  number={1},
  pages={3788},
  year={2022},
  publisher={Nature Publishing Group UK London}
}

@article{hie2024efficient,
  title={Efficient evolution of human antibodies from general protein language models},
  author={Hie, Brian L and Shanker, Varun R and Xu, Duo and Bruun, Theodora UJ and Weidenbacher, Payton A and Tang, Shaogeng and Wu, Wesley and Pak, John E and Kim, Peter S},
  journal={Nature biotechnology},
  volume={42},
  number={2},
  pages={275--283},
  year={2024},
  publisher={Nature Publishing Group US New York}
}

@article{starr2022shifting,
  title={Shifting mutational constraints in the SARS-CoV-2 receptor-binding domain during viral evolution},
  author={Starr, Tyler N and Greaney, Allison J and Hannon, William W and Loes, Andrea N and Hauser, Kevin and Dillen, Josh R and Ferri, Elena and Farrell, Ariana Ghez and Dadonaite, Bernadeta and McCallum, Matthew and others},
  journal={Science},
  volume={377},
  number={6604},
  pages={420--424},
  year={2022},
  publisher={American Association for the Advancement of Science}
}

@article{lan2020structure,
  title={Structure of the SARS-CoV-2 spike receptor-binding domain bound to the ACE2 receptor},
  author={Lan, Jun and Ge, Jiwan and Yu, Jinfang and Shan, Sisi and Zhou, Huan and Fan, Shilong and Zhang, Qi and Shi, Xuanling and Wang, Qisheng and Zhang, Linqi and others},
  journal={nature},
  volume={581},
  number={7807},
  pages={215--220},
  year={2020},
  publisher={Nature Publishing Group UK London}
}

@inproceedings{houlsby2019adapters,
  author    = {Houlsby, Neil and Giurgiu, Andrei and Jastrzebski, Stanislaw and Morrone, Bruna and {de Laroussilhe}, Quentin and Gesmundo, Andrea and Attariyan, Mona and Gelly, Sylvain},
  title     = {Parameter-efficient transfer learning for {NLP}},
  booktitle = {Proceedings of the 36th International Conference on Machine Learning},
  year      = {2019},
  series    = {Proceedings of Machine Learning Research},
  volume    = {97},
  pages     = {2790--2799}
}

@misc{skempi2dataset,
  author       = {Jankauskait{\.e}, Justina and Jim{\'e}nez-Garc{\'i}a, Brian and Dapk{\=u}nas, Justas and Fern{\'a}ndez-Recio, Juan and Moal, Iain H.},
  year         = {2019},
  title        = {{[dataset] {SKEMPI} 2.0: an updated benchmark of changes in protein--protein binding energy, kinetics and thermodynamics upon mutation}},
  howpublished = {Barcelona Supercomputing Center. \url{https://life.bsc.es/pid/skempi2/}}
}

@misc{starr2022dmsdata,
  author       = {Starr, Tyler N and Greaney, Allison J and Hannon, William W and Loes, Andrea N and Hauser, Kevin and Dillen, Josh R and Ferri, Elena and Farrell, Ariana Ghez and Dadonaite, Bernadeta and McCallum, Matthew and others},
  year         = {2022},
  title        = {{[dataset] Deep mutational scanning of the SARS-CoV-2 RBD in variant backgrounds.}},
  howpublished = {GitHub. \url{https://github.com/jbloomlab/SARS-CoV-2-RBD_DMS_variants}}
}

@misc{pdb6m0j,
  author       = {Wang, X. and Lan, J. and Ge, J. and Yu, J. and Shan, S.},
  year         = {2020},
  title        = {{[dataset] Crystal structure of {SARS-CoV-2} spike receptor-binding domain bound with {ACE2}, {PDB} entry 6M0J}},
  howpublished = {Protein Data Bank. \url{https://doi.org/10.2210/pdb6m0j/pdb}}
}

\end{document}